\documentclass[preprint,tightenlines,aps,floats,nofootinbib]{revtex4}
  
\usepackage{epsf}

\newcommand{\notE}{\ \hbox{{$E$}\kern-.60em\hbox{/}}}
\newcommand{\notp}{\ \hbox{{$p$}\kern-.43em\hbox{/}}}
\def\D0{\mbox{D\O}}

\includegraphics{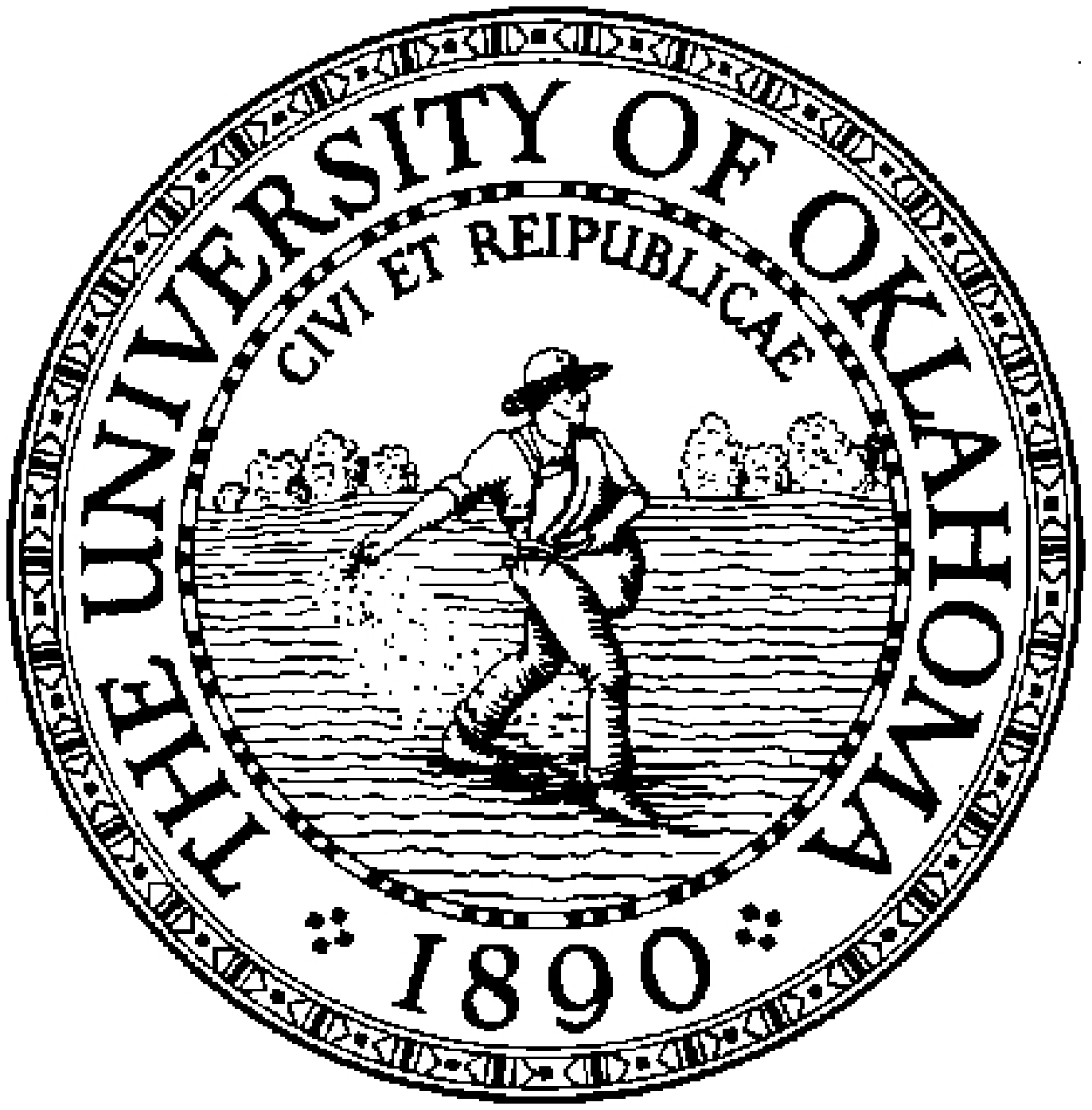}

\preprint{\font\fortssbx=cmssbx10 scaled \magstep2
\hbox to \hsize{
\hskip1.2in 
\hbox{\fortssbx The University of Oklahoma}
\hskip0.2in $\vcenter{
                      \hbox{\bf OKHEP-05-01}
                      \hbox{\bf hep-ph/0601004}
                      \hbox{December 2005}}$ }
}
  
\begin{document}  

  
\title{\vspace*{0.7in}
Detecting Higgs Bosons with Muons at Hadron Colliders}
 
\author{
Chung Kao\footnote{E-mail address: Kao@physics.ou.edu (C. Kao)} and 
Yili Wang\footnote{E-mail address: Yili@physics.ou.edu (Y. Wang)}}

\affiliation{
Homer L. Dodge Department of Physics and Astronomy \\
University of Oklahoma \\ 
Norman, Oklahoma 73019, USA 
\vspace*{.5in}}

\thispagestyle{empty}

\begin{abstract}

We investigate the prospects for the discovery of neutral Higgs
bosons with a pair of muons by direct searches at the CERN Large
Hadron Collider (LHC) as well as by indirect searches in the rare decay 
$B_s \to \mu^+\mu^-$ at the Fermilab Tevatron and the LHC.
Promising results are found for the minimal supersymmetric standard
model, the minimal supergravity (mSUGRA) model, and supergravity
models with non-universal Higgs masses (NUHM SUGRA).
For $\tan\beta \simeq 50$, we find that
(i) the contours for a branching 
fraction of $B(B_s \to \mu^+\mu^-) = 1 \times 10^{-8}$ 
in the parameter space are very close to the $5\sigma$ contours for 
$pp \to b\phi^0 \to b\mu^+\mu^- +X, \phi^0 = h^0, H^0, A^0$ at
the LHC with an integrated luminosity ($L$) of 30 fb$^{-1}$, 
(ii) the regions covered by 
$B(B_s \to \mu^+\mu^-) \ge 5\times 10^{-9}$ and the discovery
region for $b\phi^0 \to b\mu^+\mu^-$ with 300 fb$^{-1}$ are
complementary in the mSUGRA parameter space, 
(iii) in NUHM SUGRA models, a discovery of 
$B(B_s \to \mu^+\mu^-) \simeq 5\times 10^{-9}$ at the LHC will cover
regions of the parameter space beyond the direct search 
for $b\phi^0 \to b\mu^+\mu^-$ with $L = 300$ fb$^{-1}$.

\end{abstract}

%

\maketitle

\newpage

\section{Introduction}

The minimal supersymmetric standard model (MSSM) \cite{MSSM} 
has two Higgs doublets $\phi_1$ and $\phi_2$ that couple to 
fermions with weak isospin $-1/2$ and $+1/2$ respectively \cite{Guide}. 
After spontaneous symmetry breaking, there remain five physical Higgs
bosons:
a pair of singly charged Higgs bosons $H^{\pm}$,
two neutral CP-even scalars $H^0$ (heavier) and $h^0$ (lighter),
and a neutral CP-odd pseudoscalar $A^0$.
The Higgs potential is constrained by supersymmetry (SUSY)
such that all tree-level Higgs boson masses and couplings 
are determined by just two independent parameters,  
commonly chosen to be the mass of the CP-odd pseudoscalar ($m_A$) 
and the ratio of vacuum expectation values of neutral Higgs fields 
($\tan\beta \equiv v_2/v_1$). 

In the MSSM and two Higgs doublet models with Model II of Yukawa 
interactions, the couplings of down type quarks and leptons with 
the neutral Higgs bosons are proportional to $1/\cos\beta$. 
Thus a large value of $\tan\beta$ greatly enhances the production rate 
of Higgs bosons produced in association with bottom quarks
as well as the branching fraction of the rare decay 
$B_s \to \mu^+\mu^-$ mediated by neutral Higgs bosons.

If $\tan\beta \agt 10$, the MSSM neutral Higgs bosons are dominantly 
produced from bottom quark fusion $b\bar{b} \to \phi^0$
\cite{Dicus1,Dicus2,Balazs,Maltoni,Harlander} 
at the CERN Large Hadron Collider (LHC). 
For a Higgs boson produced along with one bottom quark at high
transverse momentum ($p_T$), the leading-order subprocess is
$bg \to b \phi^0$ \cite{Choudhury,Huang,Scott,Cao,Dawson:2004sh}.
If two high $p_T$ bottom quarks are required in association with
a Higgs boson, the leading order subprocess should be
$gg \to b\bar{b}\phi^0$ \cite{Dicus1,hbbmm,Plumper,Dittmaier,Dawson}.
We note that the importance of the process with a bottom quark was 
suggested by the authors of Ref. \cite{Scott}.

The LHC has a great potential to  discover the inclusive muon pair channel 
for neutral Higgs bosons of minimal supersymmetry \cite{Nikita,CMS,ATLAS}.  
Recently, it was found that the discovery channel with one energetic
bottom quark \cite{hbmm} greatly improves the discovery potential of
the LHC beyond the inclusive channel without bottom quarks
\cite{Nikita} 
($pp \to \phi^0 \to \mu^+\mu^- +X$)
and the associated channel with two bottom quarks \cite{hbbmm} 
($pp \to b\bar{b} \phi^0 \to b\bar{b} \mu^+\mu^- +X$). 
Since the ATLAS and the CMS both have excellent muon mass resolution, 
this discovery channel will provide a good opportunity to 
reconstruct the Higgs boson mass at the LHC with high precision.

We follow the strategies developed in Ref. \cite{hbmm} to investigate 
the discovery at the LHC of a neutral Higgs boson produced with one
bottom quark followed by Higgs decay into a muon pair.  
We work within the framework of the minimal supersymmetric model, 
the minimal supergravity unified model, and supergravity unified
models with non-universal Higgs boson masses at the grand unified
scale ($M_{\rm GUT}$).

In the minimal supergravity unified model \cite{mSUGRA}, 
the significance of $pp \to \phi^0 \to \mu^+\mu^-+X$ is 
greatly improved by a large $\tan\beta$ \cite{Vernon} because  
the large $b\bar{b}\phi^0$ couplings make $m_A$ and $m_H$ small
through the evolution of renormalization group equations \cite{Baer}.
Consequently, the production cross section is further enhanced 
by a large value of $\tan\beta$.

The rare decay $B_s \to \mu^+\mu^-$ has a small branching fraction 
\begin{eqnarray}
{\rm B}(B_s \to \mu^+\mu^-) = 3.4 \times 10^{-9} 
\end{eqnarray}
in the Standard Model (SM) of electroweak interactions 
\cite{Buchalla:1995vs,Buras:2003td}.
A recent calculation \cite{Krutelyov:2005tc} suggests 
\begin{eqnarray}
{\rm B}(B_s \to \mu^+\mu^-) = (5.1\pm 1.1)\times 10^{-9}
\end{eqnarray}
with updated parameters.
The current experimental upper limit is
\begin{eqnarray}
B(B_s \to \mu^+\mu^-) < 1.5 \times 10^{-7}
\end{eqnarray}
obtained by the Collider Detector at Fermilab (CDF) and the $\D0$ 
collaborations \cite{Bernhard:2005yn}.
While this branching fraction is small in the SM, it could become large 
in supersymmetric models \cite{Choudhury:1998ze}-\cite{Ellis:2005sc}
and this rare decay provides a
possible opportunity for the CDF and the \D0 experiments to discover 
new physics in the near future.

In this letter, we investigate the discovery potential of 
the direct searches for the Higgs bosons 
$pp \to b\phi^0 \to b\mu^+\mu^- +X$ at the LHC and that of 
the indirect searches for Higgs bosons in $B_s \to \mu^+\mu^-$ 
at the Fermilab Tevatron Run II within the framework of 
supersymmetric models.
We make three major contributions: (a) studying the LHC discovery potential
for $pp \to b\phi^0 \to b\mu^+\mu^- +X$ in the minimal supergravity
model (mSUGRA) and in supergravity models with non-universal Higgs bosons
masses at $M_{\rm GUT}$ (NUHM SUGRA),
(b) evaluating $B(B_s \to \mu^+\mu^-)$ in NUHM SUGRA models, and
(c) comparing these two promising channels in the MSSM, the mSUGRA, 
and non-universal SUGRA models.
In Section II, we discuss our strategies and results for the minimal 
supersymmetric standard model.
Sections III presents the discovery contours in the
parameter space of the mSUGRA model as well as that of the NUHM SUGRA
models. Conclusions are drawn in Section IV.

\section{The Minimal Supersymmetric Standard Model}

In this section, we consider the direct searches for 
$pp \to b\phi^0 \to b\mu^+\mu^- +X$ at the LHC and 
indirect searches for Higgs bosons in $B_s \to \mu^+\mu^-$ 
at the Fermilab Tevatron Run II within the framework of 
the minimal supersymmetric standard model.

\subsection{$b\phi^0 \to b\mu^+\mu^-$}

Applying previous calculations \cite{hbmm} for the Higgs signal 
at the LHC we evaluate the cross section of 
$pp \to b \phi^0 \to b \mu^+\mu^- +X$ 
with the Higgs production cross section $\sigma(pp \to b \phi^0 +X)$
multiplied by the branching fraction of the Higgs decay into muon
pairs $B(\phi^0 \to \mu^+\mu^-)$.
The cross section for $pp \to b \phi^0 +X$
($\phi^0 = H^0, h^0, A^0$) via $bg \to b \phi^0$ is calculated 
with the parton distribution functions of CTEQ6L1 \cite{CTEQ6} 
with a factorization scale $\mu_F = M_H/4$ \cite{Maltoni,Plehn}. 
Unless explicitly specified, $b$ represent a bottom quark ($b$) or 
an anti-bottom quark ($\bar{b}$).
We have also taken the renormalization scale to be $M_H/4$ that
effectively reproduces the effects of next-to-leading order (NLO) 
\cite{Scott} with a $K$ factor of one for the Higgs signal.
The bottom quark mass in the $\phi^0 b\bar{b}$ Yukawa coupling
is chosen to be the NLO running mass $m_b(\mu_R)$ \cite{bmass},
which is calculated with $m_b({\rm pole}) = 4.7$ GeV and the NLO
evolution of the strong coupling~\cite{alphas}.

In our analysis, we consider dominant physics backgrounds to the final
state of $b \mu^+\mu^-$ from $bg \to b \mu^+\mu^-$ ($b\mu\mu$)
as well as $gg \to b\bar{b}W^+W^-$ and $q\bar{q} \to b\bar{b}W^+W^-$ ($bbWW$) 
followed by the decays of $W^\pm \to \mu^\pm \nu_\mu$.
In addition, we have included the background from 
$bg \to b\mu^+\nu \mu^-\bar{\nu}$ and
$\bar{b}g \to \bar{b}\mu^-\bar{\nu} \mu^+\nu$,  
which has major contributions from 
$bg \to tW^-$ and $\bar{b}g \to \bar{t}W^+$ ($tW$). 
The muons from $b$ decays can be removed effectively with isolation cuts 
\cite{Nikita}.
We apply a K factor of 1.3 for the $b\mu\mu$ background 
\cite{Campbell}, a K factor of 2 for $bbWW$ \cite{Bonciani,Nason}, 
and a K factor of 1.5 for $tW$ \cite{Zhu}. 
Furthermore, we consider backgrounds from 
$pp \to j \mu^+\mu^- +X, j = g, q$ or $\bar{q}$ with $q = u, d, s, c$, 
where a jet is mistagged as a $b$ with a K factor of 1.3 for these processes.

We adopt the acceptance cuts as well as the $b$-tagging and mistagging
efficiencies of the ATLAS collaboration \cite{ATLAS}.
In each event, two isolated muons are required to have
$p_T(\mu) > 20$ GeV and $|\eta(\mu)| < 2.5$.

For an integrated luminosity ($L$) of 30 fb$^{-1}$,
we require (i) $p_T(b,j) > 15$ GeV, (ii) $|\eta(b,j)| < 2.5$, 
and (iii) the missing transverse energy ($\notE_T$) 
should be less than 20 GeV to reduce the background from 
from $bbWW$ and $tW$ which contains neutrinos.
The $b$-tagging efficiency ($\epsilon_b$) is taken to be $60\%$;
the probability that a $c$-jet is mistagged as a $b$-jet
($\epsilon_c$) is $10\%$ and
the probability that any other jet is mistagged as a $b$-jet
($\epsilon_j$) is taken to be $1\%$.
For $m_\phi < 100$ GeV, we change the requirement to $p_T(\mu) >$ 10
GeV for muons in both the Higgs signal and the background.

For a higher integrated luminosity of 300 fb$^{-1}$,
we require $p_T(b,j) > 30$ GeV and $\epsilon_b =50\%$.
In addition, the missing transverse energy ($\notE_T$) in each event
should be less than 40 GeV.

To study the discovery potential of 
$pp \to b \phi^0 +X \to b \mu^+\mu^- +X$ at the LHC, 
we calculate the background from the SM processes of 
$pp \to b \mu^+\mu^- +X$
in the mass window of
$m_\phi \pm \Delta M_{\mu^+\mu^-}$ where 
$\Delta M_{\mu^+\mu^-} \equiv 
1.64 [ (\Gamma_\phi/2.36)^2 +\sigma_m^2 ]^{1/2}$ \cite{ATLAS}.
$\Gamma_\phi$ is the total width of the Higgs boson,  
and $\sigma_m$ is the muon mass resolution which
we take to be $2\%$ of the Higgs boson mass \cite{ATLAS}.

\subsection{$B_s \to \mu^+\mu^-$}

In our analysis for $B_s \to \mu^+\mu^-$ within the framework of
minimal supersymmetry, we follow the approach in Refs. \cite{Babu,Tata} 
and adopt the formulas in Ref. \cite{Tata}. We make the following
choices:
(i) The matrix of the up-type Yukawa couplings is diagonal. 
(ii) The down-type Yukawa coupling matrix is $F_D = DV_{\rm CKM}^\dagger$,
where $D$ is the matrix diagonalized from $F_D$ and 
$V_{\rm CKM}$ is the Cabibbo-Kobayashi-Maskawa matrix.
(iii) We neglect the masses of the $d$ and the $s$ quarks as well as
terms that are second order or higher in the Wolfenstein parameter
$\lambda$.
(iv) At the tree level, the CKM matrix is the only source for 
flavor changing neutral current (FCNC), and 
(v) We include FCNC contributions from one-loop diagrams
involving charginos and up-type squarks as well as 
gluino and down-type squarks.

In addition, we adopt a common mass scale for supersymmetric (SUSY) particles 
and parameters $M_{\rm SUSY} = m_{\tilde{g}} = m_{\tilde{f}} = \mu = -A_f$, 
where $A_f = A_t = A_b = A_\tau$ are the trilinear couplings for the
third generation.
Two values of $M_{\rm SUSY}$ are considered: (a) $M_{\rm SUSY} = 350$ GeV 
and (b) $M_{\rm SUSY} = 1000$ GeV. 
 
In Figure 1, we present the contours for the branching fraction in the
MSSM $B(B_s \to \mu^+\mu^-) =$ 
$1.5 \times 10^{-7}$ (current experimental limit), 
$3 \times 10^{-8}$ , $1 \times 10^{-8}$, and 
$5 \times 10^{-9}$ as well as 
the discovery contours of $b\phi^0 \to b\mu^+\mu^-$ for integrated
luminosities of 30 fb$^{-1}$ and 300 fb$^{-1}$ at the LHC in the
$(m_A,\tan\beta)$ plane for two values of common masses: 
(a) $M_{\rm SUSY} = m_{\tilde{g}} = m_{\tilde{f}} =  350$ GeV, and
(b) $M_{\rm SUSY} = m_{\tilde{g}} = m_{\tilde{f}} = 1000$ GeV.


\begin{figure}[htb]
\centering\leavevmode
\epsfxsize=3in
\epsfbox{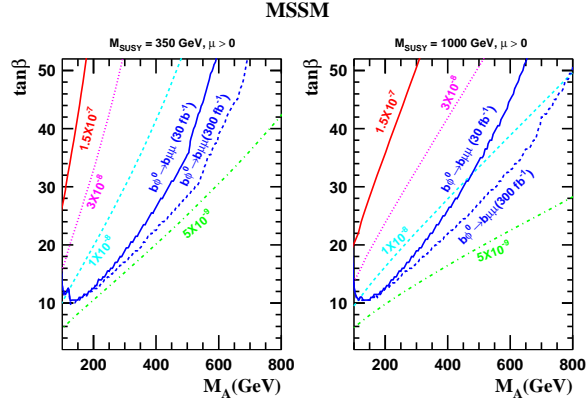}
\caption[]{
Discovery contours for $pp \to b\phi^0 \to b\mu\bar{\mu} +X$ 
at the LHC and contours of the branching of $B_s \to \mu^+\mu^-$ 
in the minimal supersymmetric standard model for 
(a) $m_{\tilde{g}} = m_{\tilde{f}} = 350$ GeV $= -A_f$ and  
(b) $m_{\tilde{g}} = m_{\tilde{f}} = 1000$ GeV $= -A_f$. 
The discovery region is the part of the parameter space above
the contour.
\label{fig:MSSM}
}\end{figure}

We note that for $M_{\rm SUSY} = 350$ GeV, the LHC will be able to
discover $pp \to b\phi^0 \to b\mu^+\mu^- +X$ with an integrated
luminosity ($L$) of 30 fb$^{-1}$ in a significantly large region of the
parameter plane beyond $B(B_s \to \mu^+\mu^-) = 3\times 10^{-8}$.
If the gluino and scalar fermions have a common
mass of approximately 1 TeV then the contour for a branching fraction of 
$B(B_s \to \mu^+\mu^-) = 1 \times 10^{-8}$ 
in the parameter plane is very close to the $5\sigma$ contour for 
$pp \to b\phi^0 \to b\mu^+\mu^- +X$ 
at the LHC with $L = 30$ fb$^{-1}$. 

Furthermore, with a higher luminosity of 300 fb$^{-1}$, the LHC will be able to
discover $pp \to b\phi^0 \to b\mu^+\mu^- +X$ for $M_{\rm SUSY} = 1000$ GeV
in a very large region of the $(m_A,\tan\beta)$ plane.
The discover contour for high luminosity with a large $M_{\rm SUSY}$ 
is very close to the contour for $B(B_s \to \mu^+\mu^-) = 5 \times 10^{-9}$
that is the not far away from the SM expectation.

\section{Supersymmetric Unified Models}

In this section, we consider both $pp \to b\phi^0 \to b\mu^+\mu^- +X$ 
and $B_s \to \mu^+\mu^-$ in the minimal supergravity model and 
supergravity models with non-universal boundary conditions for 
the Higgs boson masses at the grand unified scale ($M_{\rm GUT}$). 
We evolve supersymmetry masses and couplings from the grand unified  
scale using two-loop renormalization equations in ISAJET 7.72
\cite{ISAJET} to calculate MSSM masses, mixing angles and couplings.
The following theoretical requirements are imposed on the evolution 
of renormalization group equations:
(i) radiative electroweak symmetry breaking (EWSB) is achieved;
(ii) the correct vacuum for EWSB is obtained (tachyon free); and
(iii) the lightest SUSY particle (LSP) is the lightest neutralino 
($\chi^0_1$).

\subsection{The minimal supergravity unified model}

In the minimal supergravity (mSUGRA) model \cite{mSUGRA}, 
supersymmetry is broken in a hidden sector and SUSY breaking 
is communicated to the observable sector through gravitational interactions.  
The mSUGRA parameters are chosen to be a scalar mass ($m_0$),  
a gaugino mass ($m_{1/2}$), a trilinear coupling ($A_0$),  
the sign of a Higgs mixing parameter ($\mu$), and  
the ratio of Higgs field vacuum expectation values at the electroweak  
scale ($\tan\beta = v_2/v_1$).   
The value of $A_0$ only significantly affects results for high $\tan\beta$; 
we initially take $A_0 = 0$ and study the $A_0$ dependence later.

Figure 2 displays the discovery contours of $b\phi^0 \to b\mu^+\mu^-$ 
for an integrated luminosity of 30 fb$^{-1}$ and 300 fb$^{-1}$ at the
LHC as well as contours for four values of the branching
fraction $B(B_s \to \mu^+\mu^-)$ in the $(m_{1/2},m_0)$ plane 
of the mSUGRA model for four values of $\tan\beta =$ 20, 30, 40, and 50.
Also shown are the parts of the parameter space
(i)~disfavored by theoretical requirements
or (ii)~excluded by the chargino search at LEP 2 with 
$m_{\chi^\pm_1} < 103$ GeV.


\begin{figure}[htb]
\centering\leavevmode
\epsfxsize=3in
\epsfbox{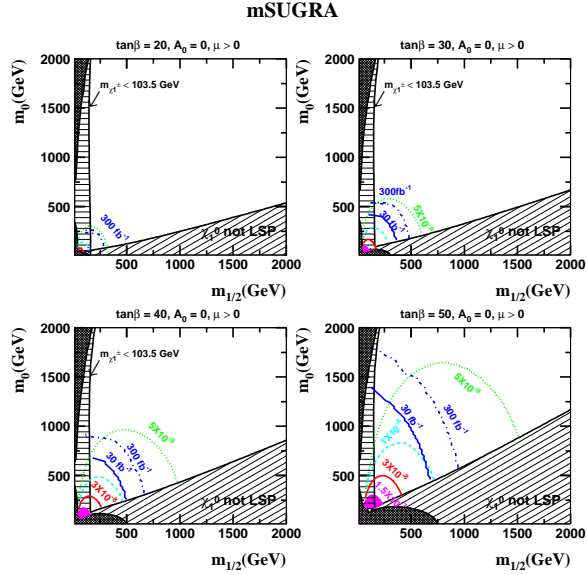}
\caption[]{
Discovery contours for $pp \to b\phi^0 \to b\mu\bar{\mu} +X$ 
at the LHC and contours of the branching of $B_s \to \mu^+\mu^-$ 
in the minimal supergravity unified model for 
(a) $\tan\beta = 20$, (b) $\tan\beta = 30$, (c) $\tan\beta = 40$, and 
(d) $\tan\beta = 50$, 
Also shown are the parts of the parameter space
(i)~excluded by theoretical requirements (slant-hatched and dark shaded),
or (ii)~excluded by the chargino search at LEP 2 (horizontally-hatched).
\label{fig:MSUGRA}
}\end{figure}

There are several interesting aspects to note in Figure 2.
\begin{itemize}
\item[(i)] If $\tan\beta \alt 30$, only a tiny region of the
 parameter space with small values of $m_{1/2}$ and $m_0$ will likely 
 lead to observable signals for either $B_s \to \mu^+\mu^-$ at
 the Tevatron Run II or $b\phi^0 \to b\mu^+\mu^-$ at the LHC.
\item[(ii)] For $\tan\beta \alt 40$,
 direct searches for $b\phi^0 \to b\mu^+\mu^-$ at the LHC with 
 $L = 30$ fb$^{-1}$ covers a much large region in the mSUGRA parameter
 space than $B(B_s \to \mu^+\mu^-) \ge 1\times 10^{-8}$.
\item[(iii)] 
If $\tan\beta \agt 50$, the discovery contour for 
$b\phi^0 \to b\mu^+\mu^-$ at the LHC with  $L = 30$ fb$^{-1}$ 
is very close to the contour for $B(B_s \to \mu^+\mu^-) \ge 1\times 10^{-8}$ 
in the mSUGRA parameter space. 
In addition, both discovery channels at the LHC become complementary. 
The direct searches for $b\phi^0 \to b\mu^+\mu^-$ 
with $L = 300$ fb$^{-1}$ covers a significant region beyond the
contour of $B(B_s \to \mu^+\mu^-) = 5\times 10^{-9}$. Likewise, the rare decay
with $B(B_s \to \mu^+\mu^-) \ge 5\times 10^{-9}$ covers a large region
beyond the discovery contour of the direct search for 
$b\phi^0 \to b\mu^+\mu^-$ with $L = 300$ fb$^{-1}$.
\end{itemize}

\subsection{The mSUGRA model with non-universal Higgs masses}

In our analysis for non-universal supergravity models, 
the GUT-scale Higgs masses are parameterized as 
\begin{eqnarray}
m_{H_i}^2 ({\rm GUT}) = (1+\delta_i) m_0^2, \;\; i = 1,2 \;.
\end{eqnarray}
The nonuniversality of Higgs-boson masses at $M_{\rm GUT}$
can significantly affect the values of Higgs masses and couplings
at the weak scale \cite{Berezinsky,nonuniversal1,nonuniversal2,munu}. 

We find that a decrease in $m_{H_1}$ with a negative $\delta_1$ 
as well as an increase in $m_{H_2}$ with a positive $\delta_2$ 
at $M_{\rm GUT}$ will lead to a smaller mass at the electroweak scale 
for the Higgs pseudoscalar ($A^0$) or the heavier Higgs scalar ($H^0$) 
than that in the mSUGRA model. Therefore, we choose three sets of
values for $\delta_i$ to study the discovery potential for detecting 
Higgs bosons with muons in SUGRA models with non-universal Higgs boson masses: 
(i) $\delta_1 = -0.5, \delta_2 = 0$,
(ii) $\delta_1 = 0, \delta_2 = 0.5$, and
(iii) $\delta_1 = -0.5, \delta_2 = 0.5$.

In Figure 3, we present the discovery contours of $b\phi^0 \to b\mu^+\mu^-$ 
for integrated luminosities of 30 fb$^{-1}$ and 300 fb$^{-1}$ at the
LHC as well as contours for four values of the branching
fraction $B(B_s \to \mu^+\mu^-)$ in the $(m_{1/2},m_0)$ plane 
for a NUHM SUGRA model with $\delta_1 = -0.5$ and $\delta_2 = 0$ 
with $\tan\beta =$ 20, 30, 40, and 50.
In addition, we show the regions of the parameter space
(i)~disfavored by theoretical requirements
or (ii)~excluded by the chargino search at LEP 2 with 
$m_{\chi^\pm_1} < 103$ GeV.


\begin{figure}[htb]
\centering\leavevmode
\epsfxsize=3in
\epsfbox{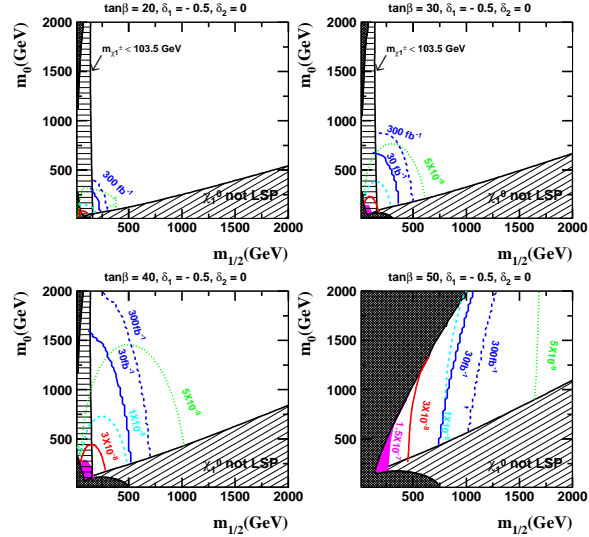}
\caption[]{
The $5\sigma$ contours for $pp \to b\phi^0 \to b\mu\bar{\mu} +X$ 
at the LHC with an integrated luminosity of 30 fb$^{-1}$ and 300 fb$^{-1}$ 
as well as contours for the branching fraction of $B_s \to \mu^+\mu^-$ 
in the ($m_{1/2},m_0$) plane of a non-universal SUGRA model
with $\mu > 0$, $A_0 = 0$ and non-universal boundary conditions
$\delta_1 = -0.5$ and $\delta_2 = 0$,
for
(a) $\tan\beta = 20$, (b) $\tan\beta = 30$, (c) $\tan\beta = 40$, and 
(d) $\tan\beta = 50$, 
Also shown are the parts of the parameter space
(i)~excluded by theoretical requirements (slant-hatched and dark shaded),
or (ii)~excluded by the chargino search at LEP 2 (horizontally-hatched).
\label{fig:NONU-SUGRA1}
}\end{figure}

Figure 4 shows contours for four values of the branching
fraction $B(B_s \to \mu^+\mu^-)$ in the $(m_{1/2},m_0)$ plane 
in a supergravity unified model with $\delta_1 = 0$ and $\delta_2 = 0.5$.
Also shown are the discovery contours of 
$pp \to b\phi^0 \to b\mu^+\mu^- +X$ 
for integrated luminosities of 30 fb$^{-1}$ and 300 fb$^{-1}$ 
at the LHC in the $(m_{1/2},m_0)$ plane
for four values of $\tan\beta =$ 20, 30, 40, and 50.


\begin{figure}[htb]
\centering\leavevmode
\epsfxsize=3in
\epsfbox{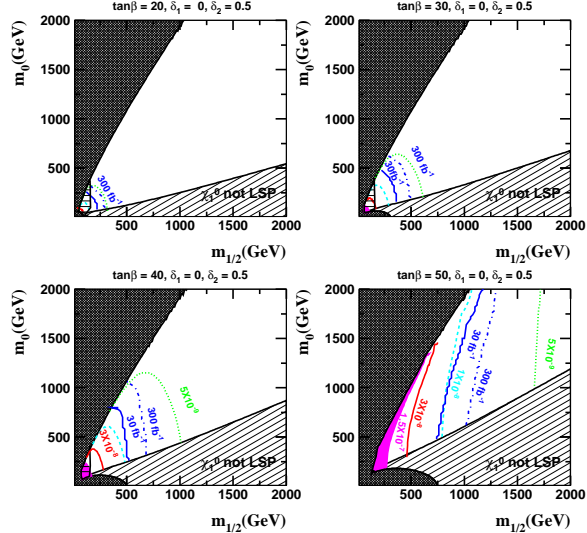}
\caption[]{
The $5\sigma$ contours for $pp \to b\phi^0 \to b\mu\bar{\mu} +X$ 
at the LHC with an integrated luminosity of 30 fb$^{-1}$ and 300 fb$^{-1}$ 
as well as contours for the branching fraction of $B_s \to \mu^+\mu^-$ 
in the ($m_{1/2},m_0$) plane of a non-universal SUGRA model
with $\mu > 0$, $A_0 = 0$ and non-universal boundary conditions
$\delta_1 = 0$ and $\delta_2 = 0.5$,
for
(a) $\tan\beta = 20$, (b) $\tan\beta = 30$, (c) $\tan\beta = 40$, and 
(d) $\tan\beta = 50$, 
Also shown are the parts of the parameter space
(i)~excluded by theoretical requirements (slant-hatched and dark shaded),
or (ii)~excluded by the chargino search at LEP 2 (horizontally-hatched).
\label{fig:NONU-SUGRA2}
}\end{figure}
%


If both Higgs boson masses are different from the common scalar mass 
at $M_{\rm GUT}$, then theoretically favored region shrinks greatly.
We present contours for four values of the branching
fraction $B(B_s \to \mu^+\mu^-)$ in the $(m_{1/2},m_0)$ plane 
in a supergravity unified model with $\delta_1 = -0.5$ and 
$\delta_2 = 0.5$ in Figure 5.
In addition, we show the discovery contours of 
$pp \to b\phi^0 \to b\mu^+\mu^- +X$ 
for integrated luminosities of 30 fb$^{-1}$ and 300 fb$^{-1}$ 
at the LHC in the $(m_{1/2},m_0)$ plane
for four values of $\tan\beta =$ 20, 30, 40, and 50.


\begin{figure}[htb]
\centering\leavevmode
\epsfxsize=3in
\epsfbox{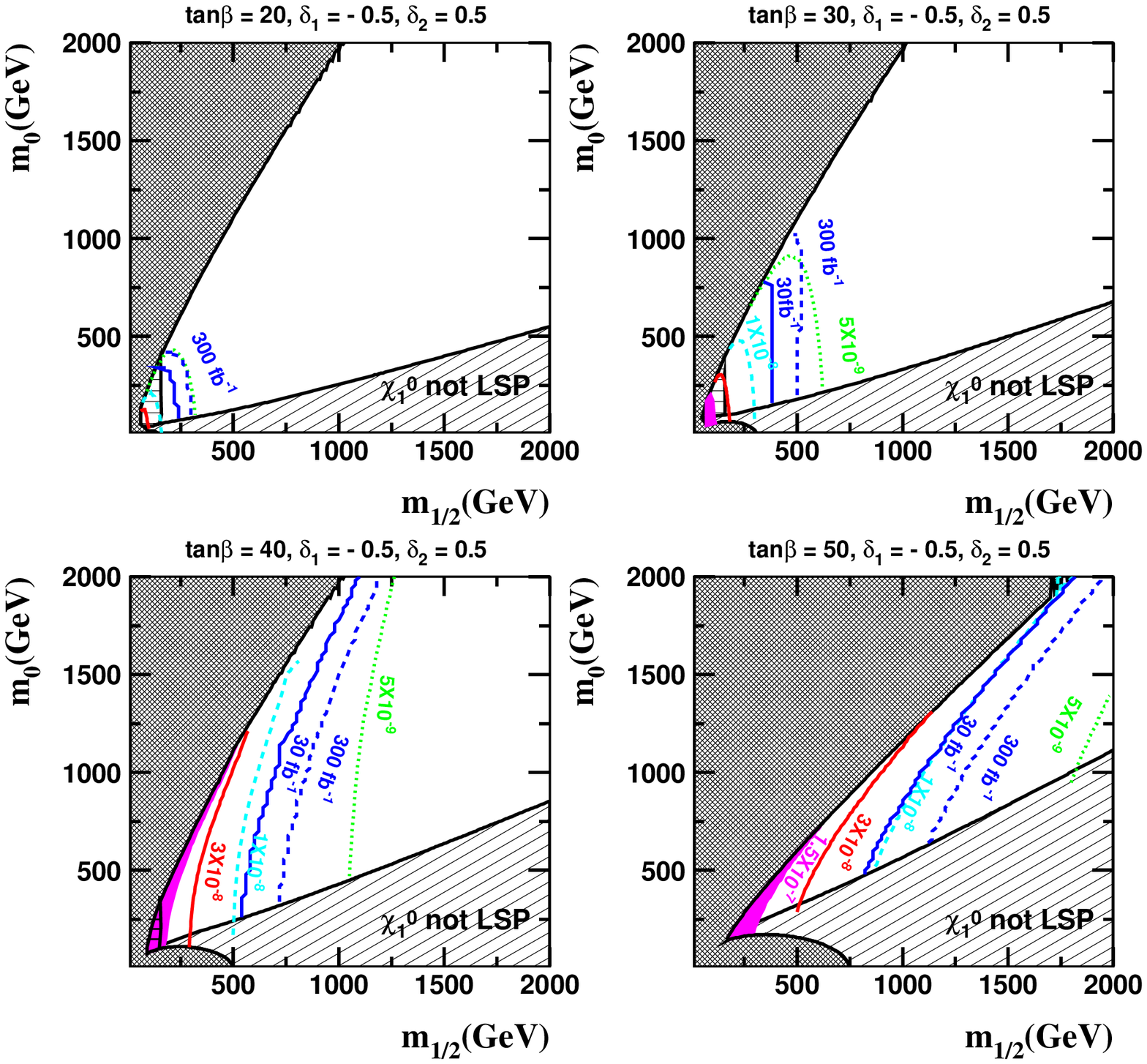}
\caption[]{
The $5\sigma$ contours for $pp \to b\phi^0 \to b\mu\bar{\mu} +X$ 
at the LHC with an integrated luminosity of 30 fb$^{-1}$ and 300 fb$^{-1}$ 
as well as contours for the branching fraction of $B_s \to \mu^+\mu^-$ 
in the ($m_{1/2},m_0$) plane of a non-universal SUGRA model
with $\mu > 0$, $A_0 = 0$ and non-universal boundary conditions
$\delta_1 =  -0.5$ and $\delta_2 = 0.5$,
for
(a) $\tan\beta = 20$, (b) $\tan\beta = 30$, (c) $\tan\beta = 40$, and 
(d) $\tan\beta = 50$, 
Also shown are the parts of the parameter space
(i)~excluded by theoretical requirements (slant-hatched and dark shaded),
or (ii)~excluded by the chargino search at LEP 2 (horizontally-hatched).
\label{fig:NONU-SUGRA3}
}\end{figure}

In all three NUHM SUGRA cases that we have considered, $m_A$ and
$m_H$ are smaller than those in the mSUGRA model for the same values
of $m_0$ and $m_{1/2}$. Consequently, both $b\phi^0 \to b\mu^+\mu^-$ 
and $B_s \to \mu^+\mu^-$ will be able to cover regions of the
parameter space with larger values of $m_0$ and $m_{1/2}$.
We note that for $\tan\beta \agt 50$, 
the observable region for $b\phi^0 \to b\mu^+\mu^-$ at the LHC with 
$L = 30$ fb$^{-1}$ is comparable to that of 
$B(B_s \to \mu^+\mu^-) \ge 1\times 10^{-8}$.
However, the contour for $B(B_s \to \mu^+\mu^-) = 5\times 10^{-9}$ lies beyond
the discovery contour for $b\phi^0 \to b\mu^+\mu^-$ at the LHC with 
$L = 300$ fb$^{-1}$.

In the NUHM SUGRA model with $\delta_1 = -0.5$ and $\delta_2 = 0$, 
($m_{H_1} = 0.707 m_0$ and $m_{H_2} = m_0$), most of the 
$(m_{1/2},m_0)$ plane is theoretically favored for $\tan\beta \alt
40$. If $m_{H_2}$ is larger than $m_0$ with $\delta_2 = 0.5$, 
the theoretically disfavored region grows rapidly as the value of
$\tan\beta$ increases.

\section{Conclusions}

In supersymmetric models, the muon pair discovery channels offer 
great promise for the detection of indirect Higgs signatures 
in $B_s \to \mu^+\mu^-$ at the Fermilab Tevatron as well as 
for the direct signal of $pp \to b\phi^0 \to b\mu^+\mu^- +X$ at the CERN LHC.

If scalar fermions and gluino have a mass close to 1000 GeV, then 
the exclusion contours of the Tevatron search for 
$B(B_s \to \mu^+\mu^-) \simeq 1\times 10^{-8}$ are comparable to the
discovery contours of the LHC search for $b\phi^0 \to b\mu^+\mu^-$ 
with an integrated luminosity of 30 fb$^{-1}$. 
However, if SUSY particles have a mass close to 350 GeV, the direct
search for $b\phi^0 \to b\mu^+\mu^-$ at the LHC becomes much more
promising than $B_s \to \mu^+\mu^-$.

In supergravity unified models, 
the branching fraction of $B_s \to \mu^+\mu^-$ and 
the significance of $pp \to b\phi^0 \to b \mu^+\mu^-+X$ 
are greatly improved by a large $\tan\beta$ because  
the large $b\bar{b}\phi^0$ couplings make $m_A$ and $m_H$ small
through the evolution of renormalization group equations and enhance
the production cross section for Higgs bosons even more.
In the mSUGRA model and in supergravity models with non-universal
Higgs masses, the direct signal of $b\phi^0 \to b\mu^+\mu^-$ at the
LHC can be discovered with a luminosity of 30 fb$^{-1}$ in a large
region of the parameter space that is comparable to 
that of $B(B_s \to \mu^+\mu^-) = 1 \times 10^{-8}$ for $\tan\beta \alt 50$.

For a large value of $\tan\beta$, the Tevatron Run II with an integrated 
luminosity of 10 fb$^{-1}$ will be able to observe 
the indirect Higgs signal with 
$B(B_s \to \mu^+\mu^-) \alt 1 \times 10^{-7}$ \cite{Dedes:2004yc} 
which then can be confirmed by the direct search of 
$pp \to b\phi^0 \to b\mu^+\mu^- +X$ at the LHC \cite{hbmm}.
Even if there are no signs of new physics at the Tevatron Run II,
we will be able to set meaningful limits on important parameters
such as the Higgs masses and
the ratio of the vacuum expectation values of Higgs fields
$\tan\beta \equiv v_2/v_1$ \cite{Kane:2003wg} for the MSSM with minimal
flavor violation.
The Tevatron Run II with an integrated luminosity of 8 fb$^{-1}$ 
will be able to exclude $B(B_s \to \mu^+\mu^-) \alt 3 \times 10^{-8}$ 
\cite{projection} which then can provide important guidance
to detect $pp \to b\phi^0 \to b\mu^+\mu^- +X$ at the LHC \cite{hbmm}.

If $\tan\beta \agt 50$, the regions covered by 
$B(B_s \to \mu^+\mu^-) \ge 5\times 10^{-9}$ and the discovery
region for $b\phi^0 \to b\mu^+\mu^-$ with 300 fb$^{-1}$ are
complementary in the mSUGRA parameter space:
the direct searches for 
$b\phi^0 \to b\mu^+\mu^-$ with $L = 300$ fb$^{-1}$ can cover 
a significantly large region beyond 
$B(B_s \to \mu^+\mu^-) \ge 5\times 10^{-9}$, and vice versa.
However, in NUHM SUGRA models, a discovery of 
$B(B_s \to \mu^+\mu^-) \simeq 5\times 10^{-9}$ at the LHC will cover
regions of the parameter space beyond the direct search 
for $b\phi^0 \to b\mu^+\mu^-$ with $L = 300$~fb$^{-1}$.

\section*{Acknowledgments}

C.K. would like to thank the Aspen Center for Physics 
where part of the research was completed, for their hospitality.
This research was supported in part by the U.S. Department of Energy
under Grants No.~DE-FG02-04ER41305 and No.~DE-FG02-03ER46040.
 

\end{document}